\documentclass[12pt]{article}
\usepackage{amsmath,amsfonts,amssymb,graphicx,epsfig}
\date{\today}
\def\unit{\leavevmode\hbox{\small1\kern-3.6pt\normalsize1}}
\newcommand{\be}{\begin{equation}}
\newcommand{\ee}{\end{equation}}
\newcommand{\bea}{\begin{eqnarray}}
\newcommand{\eea}{\end{eqnarray}}
\def\ie{{\it i.e.}}
\def \ben{\begin{enumerate}}
\def \een{\end{enumerate}}
\def \bit{\begin{itemize}}
\def \eit{\end{itemize}}
\def\ie{{\it i.e.}}
\def\lsim{\raise0.3ex\hbox{$\;<$\kern-0.75em\raise-1.1ex\hbox{$\sim\;$}}}
\def\gsim{\raise0.3ex\hbox{$\;>$\kern-0.75em\raise-1.1ex\hbox{$\sim\;$}}}

\def \av#1{\left\langle #1\right\rangle}
\def \eff{\mathrm{eff}}
\def \vckm{V_{\mathrm{CKM}}}
\def \subfermion{\mathrm{fermion}}
\def \subhiggs{\mathrm{Higgs}}
\def \sb{\mathrm{SB}}
\def \susy{\mathrm{SUSY}}
\def \GeV{{\mathrm{GeV}}}

\def \TeV{{\mathrm{TeV}}}
\def \Im{{\mathrm{Im}}\,}

\def \braket#1#2#3{\langle #1|#2| #3\rangle}

\def \cl#1{{#1\%\ \mathrm{C.L.}}}

\def \cp{\mathrm{CP}}

\def \diag{{\mathrm{diag}}}

\def \ea{{\it et al.}}
\def \eq#1{Eq.~(\ref{#1})}

\def \heff{H_{\mathrm{eff}}}
\def \hc{\mathrm{H.c.}}

\def \b{\beta}
\def \f{\phi}
\def \D{\Delta}
\def \g{\gamma}

\def \d{\delta}
\def \epsi{\epsilon}

\def \l{\lambda}

\def \m{\mu}

\def\21{$SU(2) \ot U(1)$}
\def\ot{\otimes}
\def\ie{{\it i.e.}}

\def\vev#1{\left\langle #1\right\rangle}

\def\bold#1{\setbox0=\hbox{$#1$}
     \kern-.025em\copy0\kern-\wd0
     \kern.05em\copy0\kern-\wd0
     \kern-.025em\raise.0433em\box0 }
\def \chargino{\tilde{\chi}^{\pm}}

\def \gluino{\tilde{g}}
\def \higgs{H^{\pm}}
\def \neutralino{\tilde{\chi}^0}

\def \squark{\tilde{q}}

\def \squarki#1{\tilde{q}_{#1}}

\def \wino{\tilde{W}}

\def \higgsino{\tilde{H}}
\def\arnps#1#2#3{{\it Annu. Rev. Nucl. Part. Sci.\/} {\bf#1} (#2) #3}
\def\epjc#1#2#3{{\it Eur. Phys. J.\/} {\bf C #1} (#2) #3}

\def\ibid#1#2#3{\emph{ibid.} {\bf #1} (#2) #3}

\def\npb#1#2#3{{\it Nucl.~Phys.\/}~{\bf B #1} (#2) #3}

\def\plb#1#2#3{{\it Phys.~Lett.\/}~{\bf B #1} (#2) #3}

\def\prd#1#2#3{{\it Phys.~Rev.\/}~{\bf D#1} (#2) #3}

\def\hpph#1{{\tt hep-ph/#1}}

\def\jhep#1#2#3{{\it J.~High Energy Phys.}~{\bf #1} (#2) #3}
\begin{document}

\renewcommand{\thefootnote}{\fnsymbol{footnote}}
\setcounter{footnote}{1}
\renewcommand{\baselinestretch}{1.2} \large\normalsize
\vspace{.3cm}
\begin{center}
{\bf{\Large 
Spontaneous CP violation in supersymmetric models}}\\

\vspace* {.8cm}
{\large 
Ana M. Teixeira}\\
\vspace* {5mm}
{\it Centro de F\'\i sica das Interac{\c c}{\~o}es 
Fundamentais (CFIF),
Departamento de F{\'\i}sica,  Instituto Superior T{\'e}cnico, Av.
Rovisco Pais,  1049-001 Lisboa, Portugal}\\

\vspace*{10mm}
{\bf \large Abstract}
\end{center}
{\small
We briefly comment on the question of spontaneous CP violation 
for several models of weak interactions.
We focus on one of the minimal extensions of the Standard Model where
spontaneous CP violation is viable, the next-to-minimal supersymmetric
standard model (NMSSM), with two Higgs doublets and a gauge singlet.
We analyse the most general Higgs potential without a discrete 
$Z_3$ symmetry, and derive an upper bound on the mass of the lightest
neutral Higgs boson. We estimate $\epsilon_K$ by applying the mass 
insertion approximation, finding that in order to account for the 
observed $\cp$ violation in the neutral kaon sector 
a non-trivial flavour structure in the soft-breaking $A$ terms is 
required and that the upper bound on the lightest Higgs-boson mass 
becomes stronger. We also discuss the implications of electric dipole 
moments of the electron and the neutron in SUSY models with SCPV.}

\section{\bf \large Introduction}
The origin of $\cp$ violation remains a fundamental open question in
particle physics.
In spite of the standard model success in accommodating the
experimental value of $\epsilon_K$ and 
$\epsilon'/\epsilon$ through the Cabib\-bo-Ko\-ba\-ya\-shi-Maskawa mechanism, 
an alternative scenario is to assume that rather than being explicitly 
broken at the Lagrangian level, CP is a symmetry of the Lagrangian, 
which is spontaneously broken by the vacuum~\cite{TDLee}.
We begin with a brief overview of several possible scenarios for 
spontaneous CP violation (SCPV): the standard model (SM), multi-Higgs
doublet models and the minimal supersymmetric standard model (MSSM).

As in Reference~\cite{SCPV}, we shall discuss in detail 
spontaneous CP breaking at the tree level in the context of
supersymmetry, most specifically in an extension of the MSSM with 
one gauge singlet field ($N$) besides the two Higgs
doublets ($H_{1,2}$), the so-called next-to-minimal supersymmetric 
standard model (NMSSM). In this class of models CP
violation is caused by the phases associated with the vacuum
expectation values of the Higgs fields, thus the reality of the 
CKM matrix is automatic and not an \emph{ad hoc} assumption. 
In summary, we will investigate whether or not one 
can achieve spontaneous breaking of $\cp$ whilst generating the
observed amount of $\epsilon_K$ and having Higgs-boson masses
that are consistent with experimental data.

\section{\bf \large Spontaneous CP violation: standard model and beyond}
As previously discussed, and although nearly 40 years have passed
since the first experimental evidence of CP violation,
we are still far from having a complete theoretical framework that can
fully account for both the origin of the CP symmetry break and the
measured CPV observables.

In the SM, CP is explicitly broken at the Lagrangian
level through complex Yukawa couplings. 
In the interaction basis, and prior to electroweak symmetry breaking,
the charged current and Yukawa terms in the Lagrangian can be written
as
\begin{equation}\label{SCPV:L0}
\mathcal{L}^{int} = -\bar{u^0}_L \; h_u \; \tilde{\phi} \; u_R^0
-\bar{d^0}_L \; h_d \; \phi \; d_R^0 
+ i \frac{g}{\sqrt{2}} \; \bar{u^0}_L \; \gamma^\mu \; d_L^0 \;
W_\mu^+ 
+ \hc \;,
\end{equation}
where $h_{u,d}$ are generic $3 \times 3$ complex Yukawa matrices. 
After $SU(2)_L \times U(1)_Y$ is broken to $U(1)_{em}$, CP violation
arises from the misalignment of mass and charged-current interaction
eigenstates. In the physical quark basis, the Lagrangian now reads
\begin{equation}\label{SCPV:Lphys}
\mathcal{L}^{phys} = -\bar{u}_L \; M_u^{\diag} \; u_R
-\bar{d}_L \; M_d^{\diag} \; d_R
+ i \frac{g}{\sqrt{2}} \; \bar{u}_L \; \gamma^\mu \; \vckm\; d_L \;
W_\mu^+ + \hc \;.
\end{equation}
The Cabibbo-Kobayashi-Maskawa matrix ($\vckm$) can be parametrized by 3 real
angles and one physical phase, $\delta_{\mathrm{CKM}}$, which is the
only source of CP violation in the SM. So far, the KM mechanism 
has been able to account for the experimental values of the CP 
violating parameters in the $K$ and $B$ sectors.
Nevertheless, it is not clear whether this mechanism of ``hard'' CP
violation is the only existing scenario, or if the SM contributions 
are the dominant or even the only ones.
Further motivation to discuss other scenarios of CP violation
stems from the fact that within the SM the amount of CP violation 
is not enough to generate the observed baryon asymmetry of the
Universe.

A possible way to overcome this problem is to increase the number of CP
violating sources in the theory. As an example, one can refer to
models with vector-like quarks~\cite{vector}, Left-Right symmetric
models~\cite{LR}, extensions of the SM Higgs sector,
and the minimal supersymmetric standard
model (MSSM). In fact, the MSSM adds 40 new phases,
arising from both gauge and flavour sectors, that explicitly
violate CP, and these have been the object of extensive 
studies.  

On the other hand, and instead of increasing the number of ``hard''
CPV phases, one can adopt an alternative scenario where CP is
spontaneously (or ``softly'') broken. In this framework, 
CP is originally a symmetry of the Lagrangian which is broken 
at the same time as the electroweak symmetry due to complex
scalar vacuum expectation values (VEV's)\footnote{Note that due to 
Lorentz invariance only scalar fields may have non-vanishing VEV's.}.
Invariance of $\mathcal{L}$ under a generic CP transformation requires 
that the Yukawa couplings of Eq.~(\ref{SCPV:L0}) are real (for a
complete discussion see Reference~\cite{book}).
In this case, and assuming the whole theory to be CPT invariant, 
the Poincar\'e group ($\mathcal{P}$) is also a symmetry group of the
Lagrangian.

Although it is a very appealing mechanism, spontaneous CPV is not
always viable. Let us review the status of SCPV in the SM and some of
its minimal extensions. 

\noindent{\it a) Standard model}

As we will shortly see, in the SM, and in generic $SU(2)_L \times U(1)_Y$
gauge theories with only one Higgs doublet ($\phi$), SCPV cannot occur.
The SM Higgs VEV that breaks electroweak symmetry can be generically
written as a complex space-time constant quantity, 
\begin{equation}\label{SCPV:SM:Hvev}
\av{0|\phi(x_\mu)|0}=\left(
\begin{array}{c}
0\\
v \;e^{i \alpha}
\end{array}
\right) \;.
\end{equation}
Nevertheless, this vacuum is trivially CP invariant, as one can
promptly verify by considering a generic CP transformation for the
scalar Higgs field:
\begin{equation}\label{SCPV:SM:CPtransf}
(\Bbb{CP}) \; \phi(t, \overrightarrow{r}) \; (\Bbb{CP}^\dagger) = 
e^{i \beta} \; \phi(t,- \overrightarrow{r})\;,
\end{equation}
where the phase $\beta$ that appears in the above transformation is
arbitrary. It is clear that by taking $\beta = 2 \alpha$, the vacuum
is indeed CP invariant.
To overcome this, one is naturally led to extend the Higgs content of
the model.

\noindent{\it b) Extensions of the SM: multi-Higgs doublet models}

A general problem of multi-Higgs doublet models is the existence  
of flavour-changing neutral Yukawa interactions, 
arising from the addition of scalar particles (Higgs bosons, in this case)
whose Yukawa couplings are not flavour diagonal. 
Typically, these interactions give excessive contributions to
neutral meson mixing observables, and in order to satisfy the
experimental constraints on $K^0-\bar{K^0}$, $D^0-\bar{D^0}$,
$B^0-\bar{B^0}$ mixing and some rare decays, 
one has to find a way to suppress them.
Either one assumes that the additional neutral scalars are
sufficiently heavy to decouple (with masses of order TeV), or 
the existence of some mechanism to suppress the non-diagonal couplings.
Clearly the latter is the most appealing and natural way to avoid
conflict with experiment. For the case of two Higgs doublets, and 
as originally shown in Ref.~\cite{GWP}, the only way to achieve natural flavour
conservation (NFC), is by imposing that each of the scalar doublets
only couples to either up or down type quarks. This can be achieved by
imposing a discrete or reflexion symmetry on the scalar potential, for
example a $Z_2$ symmetry. Albeit, once such a symmetry is imposed, 
SCPV cannot be obtained.

One can further increase the number of Higgs doublets: in the Branco
model~\cite{Branco:3HD}, which contains three doublets, and has NFC, 
CP can be spontaneously broken and scalar particle interactions are 
the only source of CP violation.
It is also worth mentioning that SCPV can occur in models where,
instead of just enlarging the Higgs sector, one has additional fermions or
an extended gauge sector (e.g. extended Left-Right symmetric models,
models with additional heavy exotic fermions, etc.).

\noindent{\it c) Extensions of the SM: supersymmetric models}

The MSSM appears as an appealing scenario for SCPV, since it has 
by construction two Higgs doublets and NFC is automatic. Still, it is
well known that due to supersymmetric constraints on the Higgs
potential one cannot obtain SCPV at the tree level. The reason why
this occurs is analogous to that of the non-supersymmetric two-Higgs
doublet model. 

The possibility of having radiatively induced SCPV in
the MSSM has been already ruled out for this scenario leads to the
existence of a very light Higgs boson, which is incompatible with
the present experimental data. It is therefore of
interest to consider simple extensions of the MSSM such as a model
with at least one gauge singlet field ($N$) besides the two Higgs
doublets ($H_{1,2}$), the so-called next-to-minimal supersymmetric 
standard model (NMSSM), and to ask if 
one can achieve spontaneous breaking of $\cp$ in such a class of models.

\section{\bf \large Spontaneous CPV in the NMSSM}

\subsection{The Higgs potential}
We consider the most general form of the superpotential given by 
$W~=~W_{\subfermion}~+~W_{\subhiggs}$.
In addition to the usual MSSM terms, one finds new
contributions in $W_{\subhiggs}$:
\begin{equation}\label{pot:eq:WHiggs}
W_{\subhiggs}= -\lambda \widehat N 
\widehat H_1\widehat H_2-\frac{k}{3}\, {\widehat N}^3 -r \widehat N
-\mu  \widehat H_1\widehat H_2,
\end{equation}
where $\widehat N$ is a singlet superfield.
Decomposing the SUSY soft-breaking terms as 
${\mathcal L}_{\sb}={\mathcal L}_{\sb}^{\subfermion}+
{\mathcal L}_{\sb}^{\subhiggs}$,
additional soft terms will appear in ${\mathcal L}_{\sb}^{\subhiggs}$ 
\begin{equation}\label{pot:eq:softH}
-{\mathcal L}_{\sb}^{\subhiggs}=
m_{H_i}^2 H^{a*}_i H^a_i + m_N^2 N^* N
-\left(B\mu \varepsilon_{ab}H_1^a H_2^b +A_{\lambda} N \varepsilon_{ab} 
H_1^a H_2^b +\frac{A_k}{3}N^3 +A_r N + \hc \right).
\end{equation}
In the above equations $i,j=1,2,3$ denote generation indices, 
$a,b=1,2$ are SU(2) 
indices, and $\varepsilon$ is a completely antisymmetric $2\times2$  
matrix with $\varepsilon_{12}=1$. In the above expression, 
$\widehat H_1^a$ and $\widehat H_2^a$
denote the Higgs doublets of the minimal supersymmetric standard model
and $\widehat N$ is a singlet field. 
The matrices $h_U,h_D$, and $h_E$ give rise to the usual 
Yukawa interactions which generate the masses of quarks and leptons.
As pointed out before, and since in models of SCPV CP is conserved at
the Lagrangian level, these matrices are real.

In order to solve the so-called `$\mu$ problem' of the MSSM, a
discrete $Z_3$ symmetry was originally imposed on the superpotential,
which naturally leads to $\mu=r=0$. Nevertheless, in this case the 
NMSSM has no spontaneous $\cp$ violation~\cite{Romao:nogo}. 
As it can be seen from Eq.~(\ref{pot:eq:WHiggs})
we do not require the superpotential
to be invariant under a discrete $Z_3$ symmetry.
In our analysis we do not relate 
the soft SUSY-breaking parameters to some common unification scale, 
but rather take them as arbitrary at the electroweak scale.
In what follows we shall assume that the tree-level 
potential is $\cp$ conserving and set all parameters (soft squark and
Higgs masses, bilinear and trilinear soft breaking terms) to be real, 
but allow complex vacuum expectation values for the neutral Higgs
fields which emerge after spontaneous symmetry breaking: 
\begin{equation}
\vev{H_i^0}=v_i e^{i \theta_i}/\sqrt{2} \;;\quad \quad
\vev{N}=v_3 e^{i \theta_3}/\sqrt{2} \;.
\end{equation}
After deriving the $\cp$-invariant neutral scalar potential, 
we find that only the following phase combinations appear:
$\phi_D=\theta_1+\theta_2,\; \phi_N=\theta_3$. Hence, and without loss
of generality we shall set $\theta_1=0$.
\begin{figure}
\begin{center}
\epsfig{figure=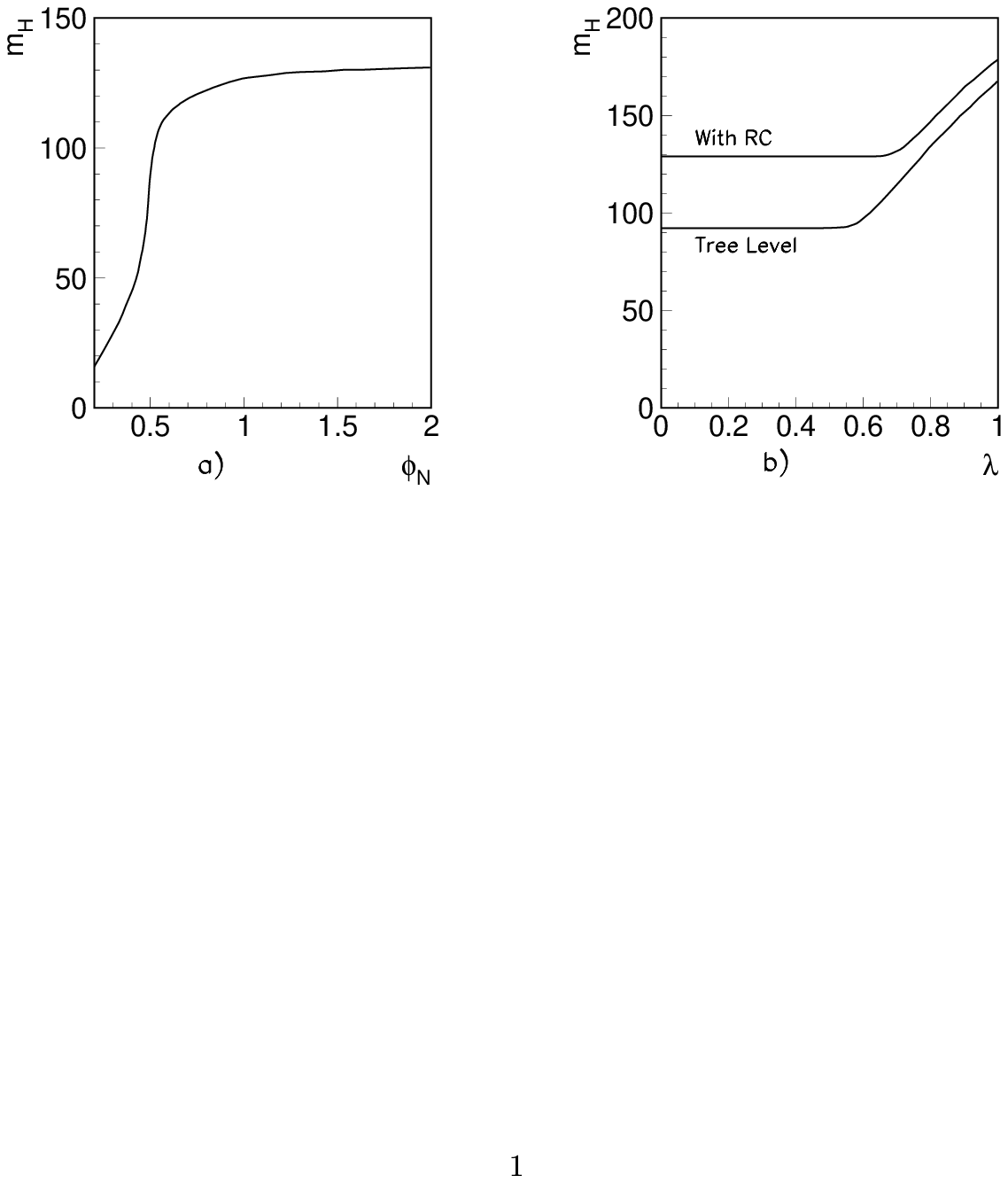,height=2.3in}
\caption{Maximum value of the lightest Higgs-boson mass (in GeV) as a
function of the $\cp$-violating phase $\phi_N$ (in radians) (a), 
and as a function of the singlet coupling $\lambda$ at the tree level 
and after including radiative corrections (at one-loop level) for 
$M_{\susy}=1\,\TeV$ (b).\label{scpv_plot_mhtheta}}
\end{center}
\end{figure}
We have found that an acceptable mass spectrum can be easily obtained with  
the exact values depending on the set of parameters we choose. 
As it can be seen in Figure~\ref{scpv_plot_mhtheta}~a), not only the
large singlet phase solution is allowed, but it is indeed favoured. 
The maximal possible value of the Higgs-boson
mass can differ from that of the MSSM for the case of large values of 
the coupling constant $\lambda$ as depicted in 
Figure~\ref{scpv_plot_mhtheta}~b).
For low values of $\lambda$, corrections to the tree level Higgs-boson
mass are significant and depend mainly on the SUSY scale that we take for
the squarks, with $max(m_{H^0})$ ranging from $105$ to $130\; \GeV$, as the
typical SUSY scale varies from $300$ to $1000 \;\GeV$.
Finally, we point out that the SM and MSSM Higgs boson mass limits
obtained at LEP do not necessarily apply to the NMSSM (see, 
e.g. ~\cite{Higgs:mass}) since due to some singlet admixture the
lightest neutral Higgs boson may have a reduced coupling to the $Z^0$
~\cite{model:nmssm:II} and thus even escape detection.

\subsection{Brief overview of the model}
In the scenario we are considering, CP invariance is imposed on the
Lagrangian, and hence all couplings are real. In particular, the
Yukawa couplings in Eq.(\ref{SCPV:L0}) are arbitrary real matrices 
in flavour space. After spontaneous symmetry breaking, 
the up and down quarks acquire masses,
\begin{equation}\label{vckm:eq:qmass}
m_U = h_U \frac{v_2}{\sqrt{2}} e^{-i\phi_D}, \quad 
m_D= h_D \frac{v_1}{\sqrt{2}}. 
\end{equation}
As it can be seen from the above equation, 
the overall phase $\phi_D$ can be rotated away by means of
a phase transformation on $u_R$, i.e. $u_R\to u_R' = e^{-i\phi_D} u_R$. 
Recalling that the $W$-boson interactions are purely left-handed, 
this phase does not appear in charged weak interactions.
Therefore, in this model the CKM matrix is real and there is no CP
violation stemming from the SM through the CKM mechanism. This is a
consequence of having real Yukawa couplings and overall phases that
can be reabsorbed in the right-handed fields (which are not involved in
$W^\pm$ boson exchange).
As we will soon discuss, CP violation will arise solely from 
the relative phases in the VEV's of the neutral Higgs fields, 
$\f_{D}$ and $\f_{N}$, which appear in the scalar quark, gaugino 
and Higgsino mass matrices, as well as in some of the vertices.

In the squark sector, working in the `super-CKM' basis, we find
complex contributions to the squark mass terms, which can be
generically written as 
\begin{equation}\label{mass:squark:gen}
M^2_{\squark}=
\left(\begin{array}{cc}
M^2_{\squarki{LL}}&M^2_{\squarki{LR}}\\
M^2_{\squarki{RL}} & M^2_{\squarki{RR}}
\end{array}\right),\quad \tilde{q}=\tilde{U}, \tilde{D}\;.
\end{equation}
In particular, we focus on the $LR$ submatrices of the up and 
down squark squared masses:
\begin{equation}\label{mass:squark:detail}
M^2_{\tilde{U}_{LR}} =M^{2\dagger}_{\tilde{U}_{RL}}=
V_L^U Y_U^*V_R^{U\dagger}\frac{v_2}{\sqrt{2}} 
- \m_{\eff} \cot\b e^{i \f_D} m_U^{\diag}\;; \; (U\rightarrow D),
\end{equation}
where $Y_U^{ij}\equiv A_U^{ij} h_q^{ij}, \;
(\mathrm{no\ sum\ over\ }i,j), \;
\m_{\eff}\equiv \m +\l  \frac{v_3}{\sqrt{2}}e^{i\f_N}.$
As we shall see in the next section, a non-universal flavour structure in the 
$A$ terms, i.e. $A^{ij}_q\neq \mathrm{constant}$, is indispensable
for having sizable supersymmetry contributions to $\cp$ violation in the 
kaon sector.

In the chargino sector (defining $m_{ \tilde{W}}=M_2, \;m_{ \tilde{H}}=
|\m_{\eff}|,\  \mathrm{and} \;\varphi= \arg \left( \m_{\eff}\right)$)
the following weak basis interaction Lagrangian arises:  
\begin{equation}\label{int:lagrangian}
-{\mathcal L}_{\mathrm{int}}= m_{ \tilde{W}}\overline{\tilde{W}}\tilde{W}
+ m_{ \tilde{H}} \overline{\tilde{H}} \tilde{H}
+\frac{g}{\sqrt{2}}(v_1 e^{-i \varphi} \overline{\tilde{W}}_R\tilde{H}_L+
v_2 e^{i \f_D} \overline{\tilde{W}}_L\tilde{H}_R+ \hc).
\end{equation}

\section{\bf \large Implications of indirect CP violation for the NMSSM}
\begin{figure}
\begin{center}
\epsfig{figure=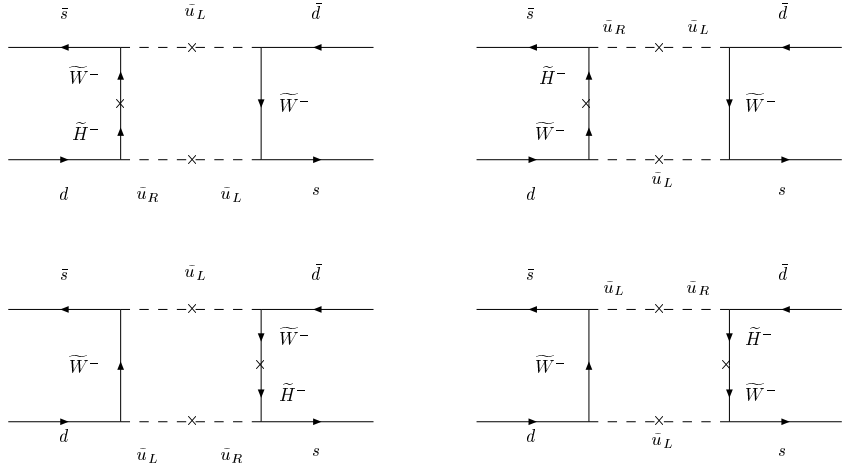,height=2.5in}
\caption{The main contributions to $\epsi_K$ in the mass 
insertion approximation with $W$-ino and Higgsino exchange.
\label{mia:leading}}
\end{center}
\end{figure}

The next step in the discussion of the viability of SCPV in the NMSSM
consists in addressing whether or not one can have CP violation in 
$K^0$--$\bar{K}^0$ mixing, and the possible implications on the upper
bound of the lightest Higgs-boson mass.
Accordingly, we will compute the box-diagram 
contributions to $\epsilon_K$ by applying the mass insertion
approximation. The effective Hamiltonian 
governing $\D S=2$ transitions can be written as 
$\heff= \sum_i c_i {\mathcal O}_i$, and the CP observable 
$\epsilon_K$ is then given by:
\begin{equation}
\epsilon_K \simeq \frac{e^{i\pi/4}}{\sqrt{2}}
\frac{\Im\mathcal{M}_{12}}{\Delta m_K}\;, \quad \text{where}\quad \quad 
\mathcal{M}_{12}=\frac{\braket{K^0}{\heff}{\bar{K}^0}}{2m_K}\;.
\end{equation}
In the presence of SUSY contributions the Wilson coefficients $c_i$ can be
decomposed as $c_i= c_i^W + c_i^{\higgs}+ c_i^{\chargino} + c_i^{\gluino} + 
c_i^{\neutralino}$. 
Recalling that the $\vckm$ matrix is real one has no $W$ boson or
charged Higgs contributions. Regarding gaugino mediated diagrams,   
and in the approximation of retaining only a 
single mass insertion in an internal squark line, we find that in the
present scenario with low $\tan\b$, we have a $c_i^{\chargino}$
dominance. As for the local operators 
${\mathcal O}_i$ \cite{fcnc:susy:constraints}, the $\D S=2$ 
transition is largely governed by the $V$--$A$ four-fermion operator 
${\mathcal O}_1 = \overline{d} \g^\m P_L s \overline{d} \g_\m P_L s$.
Therefore, we consider only the non-standard contributions to the
Wilson coefficient $c_1$, which are dominated by the diagrams depicted
in Figure~\ref{mia:leading}.
In the limit of degenerate left-handed up-type squarks, keeping only
leading top-quark contributions and using the orthogonality of the $\vckm$,
we find that the imaginary part of the neutral
kaon mass matrix off-diagonal element is
\begin{equation}\label{eK:eq:s4s13}
\Im \mathcal{M}_{12}=
\frac{2 G_F^2 f^2_K m_K m_W^4 }{3\pi^2\av{m_{\tilde{q}}}^8}
(V_{td}^*V_{ts}^{}) m_t^2 
\left|e^{i \phi_D}m_{{\wino}}+\cot\b m_{{\higgsino}}\right| 
\D A_U \sin(\varphi_{\chi}-\f_D)\;
{(M^2_{\tilde{Q}})}_{12}\;I_L \;.
\end{equation}
In the above formula, $I_L$ is the loop function (see Ref.~\cite{SCPV}) and 
$\D A_U\equiv A_U^{13}-A_U^{23}$. 
From inspection of \eq{eK:eq:s4s13}, it is 
straightforward to conclude that in order to get a non-vanishing 
$\Im \mathcal{M}_{12}$ we need a theory of non-universal $A_U$ terms
(i.e. $\D A_U\neq 0$); \ie, it is not possible to saturate the 
observed $\cp$ violation in the $K$-meson system in the context of SUSY 
with a real CKM matrix and universal $A_U$ terms. 
In Table \ref{table:res} we present the results 
\footnote{For our numerical calculations, we have used the nominal values
${(M^2_{\tilde{Q}})}_{12}/\av{m_{\tilde{q}}}^2=0.08 ,\; \\V_{ts}=-0.04,\;  
V_{td}=0.0066,\; m_t=175\,\GeV$ and $\D A_{U}=500\,\GeV$.}
for the absolute value of $\epsilon_K$ for various sets of SUSY
parameters and low $\tan\b$.
{\small
\begin{table}[ht]
\begin{center} {\footnotesize
\begin{tabular}{cccccccccc} \hline\hline
$|\epsi_K|$ &$\phi_D$  &$\phi_N$ & 
$m_{H^0}$ & $\av{m_{\tilde{q}}}$ &
$m_{\tilde{t}_R}$ &$\tan \beta$  &$\lambda$ & $v_3$\\
$(10^{-3})$&$(\mathrm{rad})$ & $(\mathrm{rad})$ &$(\GeV)$&$(\GeV)$&$(\GeV)$&
&&$(\GeV)$ \\ \hline 
$3.24$ &$4.71$   &$1.57$   &$99$ &$252$ &$235$ &$6.7$&$-0.03$ &$327 $ \\ 
$3.03$ &$0.89$   &$1.75$   &$97$ &$261$ &$168$ &$6.6$&$+0.33$ &$387$ \\ 
$2.75$ &$4.71$   &$4.71$   &$99$ &$232$ &$201$ &$9.2$&$-0.02$ &$221$ \\ 
$2.42$ &$1.96$   &$4.08$   &$94$ &$299$ &$174$ &$5.1$&$-0.06$ &$352$ \\ 
$2.10$ &$4.67$   &$4.75$   &$98$ &$279$ &$220$ &$7.8$&$+0.01$ &$142$ \\ 
$2.02$ &$4.68$   &$4.71$   &$92$ &$250$ &$152$ &$7.4$&$+0.02$ &$371$\\ 
$2.01$ &$4.18$   &$4.73$   &$96$ &$280$ &$232$ &$4.6$&$-0.01$ &$238$ \\
$1.31$ &$1.12$   &$4.72$   &$100$&$273$ &$241$ &$9.6$&$-0.01$ &$238$ \\ 
$1.29$ &$2.35$   &$4.70$   &$99$ &$258$ &$230$ &$6.1$&$-0.13$ &$363$ \\ 
\hline\hline
\end{tabular} }
\end{center}
\caption{Numerical values of $|\epsi_K|$ in the low $\tan\b$ region
for certain sets of model parameters that satisfy the minimisation 
condition of the Higgs potential.}\label{table:res}
\end{table}
}

In order to saturate the observed value of 
$|\epsilon_K|$ \cite{PDG} and to obey present experimental 
limits on the sparticle spectrum, one has to take $\D A_{U}$ of order 
$500\,\GeV$. Values of $A_U^{i3}$ ($i=1,2$) around the $\TeV$ scale do not 
significantly affect the mass spectrum of the theory, and can 
account for values of the left-right mass insertions
$(\d_{LR}^U)_{i3}$  which are consistent with present experimental
bounds~\cite{janusz}.

The large CP phases appearing in Table \ref{table:res} hint
to potential problems with the electric dipole moments (EDM's) 
of the electron and neutron. 
From the computation of the contributions to the EDM's of electron and 
neutron mediated by photino and gluino 
for the sets of parameters displayed in Table \ref{table:res}, 
we find that compatibility with the present experimental results of 
$d_n < 6.3\times 10^{-26}\, e\, \mathrm{cm}$ ($\cl{90}$) and
$d_e = 1.8\times 10^{-27}\, e\, \mathrm{cm}$ \cite{PDG},
requires that the photino and gluino masses should satisfy 
$0.5 \, \TeV \lesssim m_{\tilde{\g}} \lesssim 2\, \TeV$ and
$2  \, \TeV \lesssim m_{\gluino} \lesssim 6\, \TeV$.
Such a hierarchy in the soft gaugino masses appears to be
rather unnatural since the masses of the squarks and $W$-ino 
are typically of the order $100$--$300\, \GeV$ in this
model. Moreover, masses of the superpartners of about $1\, \TeV$ 
may be in conflict with the cosmological relic density.
Finally, one should point out 
that the above-mentioned hierarchy for the spartners leads to an
unacceptable scenario for the lightest supersymmetric particle (LSP). In this 
case, the LSP would be either charged or have a non-zero lepton number.

\section{\bf \large Conclusions}
In this talk we have addressed the viability of spontaneous CP
violation in the context of some minimal extensions of the SM.
In particular, we have studied spontaneous $\cp$ violation in the 
NMSSM, demonstrating that it is possible to generate sufficient 
$\cp$ violation in order to account for the magnitude of $\epsilon_K$.
We have shown that the minimisation of the most general 
Higgs potential leads to an acceptable mass spectrum which is accompanied by
large $\cp$-violating phases. 
Regarding $\cp$ violation in $K^0$--$\bar{K}^0$ mixing
we have discussed that saturating $\epsilon_K$ requires a 
rather low SUSY scale with $M_{\susy}\approx 300\,\GeV$ 
(i.e.~light squark and $W$-ino masses) 
and a non-trivial flavour structure of the soft SUSY-breaking trilinear 
couplings $A_U^{i3}$ ($i=1,2$). As a consequence, the
parameter space is severely constrained and the mass of 
the lightest Higgs boson is further diminished,
and it turns out to be no greater than $\sim 100\,\GeV$ for the case 
of low $\tan\b$ ($\lesssim 10$).
We have also argued that the large phase solution presents a potential
conflict with the severe constraints on the EDM's of electron and
neutron. 
Therefore, the implications of the 
EDM bounds on the parameter space, together with the implied LSP scenario,
will be a great challenge for SUSY models with spontaneous CP
violation (at least for minimal models as the one here discussed).\\

{\noindent {\bf \large Acknowlegdments}}\\
We are deeply thankful to the organizers for their kind hospitality
during the Conference.~We would also like to thank F.R. Joaquim for his help in
preparing this presentation.

\end{document}